\documentclass[submission,copyright,creativecommons]{eptcs}
\providecommand{\event}{HSB 2013} % Name of the event you are submitting to
\usepackage{color}
\usepackage{array}
\usepackage{amsmath,amssymb}
\usepackage{graphicx}
\usepackage{subfigure}

\usepackage{tikz}
\usetikzlibrary{arrows,positioning,automata}

\newtheorem{thm}{Theorem}

\newtheorem{defn}[thm]{\ \\Definition}
\newtheorem{definition}[thm]{\ \\Definition}

\renewcommand{\vec}[1]{\mathbf{#1}}
\renewcommand{\cal}[1]{\mathcal{#1}}
\newcommand{\bb}[1]{\mathbb{#1}}
\renewcommand{\phi}{\varphi}

\newcommand{\change}[1]{#1}

\title{On the Robustness of Temporal Properties\\ for Stochastic Models}
\author{Ezio Bartocci
\institute{TU Wien, Austria}
\email{ezio.bartocci@tuwien.ac.at}
\and
Luca Bortolussi
\institute{DMG, University of Trieste, Italy}
\institute{CNR/ISTI, Pisa, Italy}
\email{luca@dmi.units.it}
\and
 Laura Nenzi
\institute{IMT, Lucca, Italy}
\email{laura.nenzi@imtlucca.it}
\and
Guido Sanguinetti
\institute{School of Informatics, University of Edinburgh}
\institute{SynthSys, Centre for Synthetic and Systems Biology, University of Edinburgh}
\email{ g.sanguinetti@ed.ac.uk}
}
\def\titlerunning{On the Robustness of Temporal Properties for Stochastic Models}
\def\authorrunning{E. Bartocci, L. Bortolussi, L. Nenzi \& G. Sanguinetti}
\begin{document}
\maketitle

\begin{abstract}

Stochastic models such as Continuous-Time Markov Chains (CTMC)
and Stochastic Hybrid Automata (SHA) are powerful
 formalisms to model and to reason about the dynamics of biological systems, 
due to their ability to capture the stochasticity inherent in biological processes. A classical question in formal 
modelling with clear relevance to biological modelling is the model checking problem, i.e. calculate the probability  that a behaviour, expressed for instance in terms
of a certain temporal logic formula, may occur in a given stochastic process.
However, one may not only be interested in the notion
of satisfiability, but also in the capacity of a system 
to mantain a particular emergent behaviour unaffected by the perturbations, 
caused e.g. from extrinsic noise, or by possible small changes in the model parameters. 
To address this issue, researchers from the verification community
have recently proposed several notions of \emph{robustness} for temporal logic
providing suitable definitions of distance between a trajectory of a (deterministic) dynamical system and 
the boundaries of the set of trajectories satisfying  the property of interest. %\red{Giovani, qui mi sfugge il filo logico: diciamo che nei casi stocastici la nozione di soddisfacibilita` non basta ergo la gente ha introdotto la nozione di robustezza. Poi pero` diciamo che la nozione di robustezza la introduciamo noi in questo lavoro per i sistemi stocastici. Forse basta rimuovere ``when we deal with stochastic models'' dalla frase precedente.}
The contributions of this paper are twofold. First, we extend the  notion of  
robustness to stochastic systems, showing that this naturally leads to a distribution of robustness scores. 
By discussing two examples, we show how to  approximate the distribution of the 
robustness score and its key indicators:
the \emph{average robustness} and the \emph{conditional average robustness}.
Secondly, we show how to combine these indicators with the satisfaction
probability to address the \emph{system design problem}, where the goal is to 
optimize some control parameters of a stochastic model in order to best maximize 
robustness of the desired specifications.

%Stochastic time trace for the Repressilator
%system of Table 2. Parameters are kp = 1, kd = 0:01,
%kb = 1, ku = 0:01

\end{abstract}

% !TEX root =  main.tex

\section{Introduction}
\label{sec:introduction}

Biological systems at the single cell level are inherently stochastic. 
Molecules inside the cells  perform random movements (\emph{random walk})
and the reactions among them may occur when the probability of collision  
is high enough. The number of molecules of each species at each time point is therefore a random process: assuming instantaneous reactions, this process can be modelled as a Markovian (i.e. memoriless) discrete state, continuous time process.  When the number of molecules of each species involved is large, so that many reactions happen in any small interval of time, stochastic effects 
can be neglected. However, if the concentration of the molecules (of at least some of the species)
is low the stochasticity plays an important role and must be taken into account.
For this reason, stochastic models such as Continuous-Time Markov Chains (CTMC)~\cite{Durett2012}
and Stochastic Hybrid Automata~\cite{Bujorianu2005} are particularly powerful 
and suitable formalisms to model and to reason about biological systems 
defined as stochastic systems over time.

A classical question in formal modelling is to calculate the probability that a  
behaviour, expressed in terms of a certain temporal logic formula, may occur
in  a given stochastic process, with specified parameters. \emph{Probabilistic Model Checking}~\cite{Baier1997,Baier2003}
(PMC) is a well-established verification technique that provides a quantitative answer
to such a question. 
The algorithm used to calculate this probability~\cite{Kwiatkowska2004} produces the 
exact solution, as it operates directly on the structure of the Markov chain. Despite the success 
and the importance of PMC, this technique suffers some computational 
 limitations, either due to state space explosion or to the difficulty  (impossibility) 
in checking analytically formulae in specific logics, like Metric Temporal Logic (MTL)~\cite{Baier2003,Chen2011a}. Furthermore, PMC 
provides only a quantitative measure of
the \emph{satisfability} (yes/no answer) of a temporal logic specification (i.e., the probability of the property being true).

However, especially when we deal  with stochastic models, the notion of satisfability
may be not enough to determine the capacity of a system to mantain a particular emergent
behaviour unaffected  by the uncertainty of the perturbations due to its stochastic nature or by 
possible small changes in the model parameters.
A similar issue also arises when considering the satisfability of a property by deterministic dynamical systems which may be subject to extrinsic noise or uncertainty in the parameter. To address this question in the deterministic case, researchers from the verification community
 have proposed several notions of \emph{temporal logic based robustness}~\cite{Donze2010,Fainekos2009,Rizk2008},
providing suitable definitions of distance between a trajectory of a system and the behavioural 
property of interest, expressed in terms of a temporal logic formula. These effectively endow the logic of interest with quantitative semantics, allowing us to capture not only whether a property is satisfied but also {\it how much} it is satisfied.
A similar notion of robustness for stochastic models would clearly be desirable but, to our knowledge, has not been formalised yet.

The contributions of this paper are twofold. First,we provide a simulation-based method to define a notion of robust satisfability in stochastic models. Simulation-based approaches, such as statistical model checking~\cite{Younes2004}, 
can be be used to estimate for a stochastic model the \emph{robust satisfability distribution}
 for a given temporal logic formula, with a guarantee of asymptotic correctness. 
% : these methods are 
%usually asymptotically exact, in the limit when the number
%of simulations used is large; nevertheless, establishing what is a sufficiently large
%number of simulations to achieve a certain accuracy is a nontrivial problem~\cite{Zuliani2012}.
This distribution is the key to understand  how the behaviour specified 
by the logic temporal formula is unaffected by the stochasticity of the system.
In particular, in this paper we consider two important indicators of this distribution:
the \emph{robustness average} and the \emph{conditional robustness average} on 
a formula being true or false.  We discuss how to compute
the robust satisfability distribution and its indicators on two biological examples.
Second, we show how to combine these indicators with the satisfaction probability
to address the \emph{system design problem}, where the goal is to optimize (few) 
control parameters of a stochastic model in order to best maximize these three indicators.
The proposed approach takes advantage of Gaussian Process Upper Confidence 
Bound (GP-UCB) algorithm introduced in~\cite{gpucb}.

 The paper is structured as follows: in Section \ref{sec:basics}
we introduce the background material. In Section~\ref{sec:robustness} we
discuss the  robustness of stochastic models using the quantitative 
semantics of the Signal Temporal Logic (STL).  In Section~\ref{sec:casestudy} 
we present some experimental results for the robustness of STL
formulae for two stochastic models that we have chosen as our case
studies: the Schl\"ogl system and the Reprissilator. In Section~\ref{sec:design} 
we show an application of the robust semantics to the system design 
problem. The related works and the final discussion are in Section~\ref{sec:conclusion}.
% !TEX root =  main.tex

\section{Background}
\label{sec:basics}

%{\bf Introduce the model of genetic networks we use. Quickly.
%
%
%Introduce STL and QMiTL. Introduce robustness analysis and Breach.}  

\paragraph{Markov Population Models.}
The simplest class of stochastic processes we will consider are Continuous Time Markov Chains (CTMC) \cite{Durett2012} that describe population processes (PCTMC). A population process intuitively is a  system in which agents or objects of different kinds, and with different internal states, interact together. The classical example are biochemical and genetic networks, but other population processes include  ecological systems, computer networks, and social systems. 
%We are interested in investigating the collective behaviour of such systems.

We will describe  PCTMC by the simple formalism of biochemical reaction networks. The state of the system is described by a vector $\vec{X} = (X_1,\ldots,X_n)$ of $n$ integer-valued variables $X_i$, each counting the number of entities of a given class or species. 
The dynamics of this system is specified by a set of $m$ reactions $\mathcal{R} = \{\eta_1,\ldots,\eta_m\}$, which can be seen as description of events changing the state of the system. Each reaction $\eta_l$ is of the form
\[ r_1 X_{i_1} + \ldots + r_k X_{i_k} \rightarrow s_1 X_{j_1} + \ldots + s_h X_{j_h},\]
where $X_{i_a}$ is a reactant and $X_{j_b}$ is a product (they are both variables of $\vec{X}$), and $r_i$, $s_j$ are the stoichiometric coefficients, i.e. the amount of agents/ entities consumed or produced by the reaction. Stoichiometric information of a reaction $\eta_l$ can be condensed into an update vector $\vec{v}_l$, giving the net change in population variables due to $\eta_l$:  $\vec{v}_l = \sum_{b\leq h} s_b \vec{1}_{j_b}  - \sum_{a\leq k} r_a \vec{1}_{i_a}$, where $\vec{1}_j$ equals one in position $j$ and zero elsewhere.
Additionally, each reaction $\eta_j$ has an associated rate function $f_j(\vec{X})$ giving the rate of the transition as a function of the global state of the system. 

From a set of reactions $\mathcal{R}$ and species $\vec{X}$, we can easily derive the formal representation of a CTMC in terms of its infinitesimal generator matrix, see for instance \cite{tutorial}. Here we just recall that the state space of the CTMC is $\mathbb{N}^n$ (or a proper subset, if any conservation law is in force). Such CTMC can be simulated with standard algorithms, like SSA \cite{gillespie}.

\paragraph*{Fluid Approximation}
From a Markov population model, we can easily construct an alternative semantics in terms of Ordinary Differential Equations (ODE), assuming variables $\vec{X}$ to be continuous and interpreting each rate as a flow, thus obtaining the vector field 
\[ F(\vec{X}) = \sum_{\eta_l \in \mathcal{R}} \vec{v}_l f_l(\vec{X}),\]
defining the ODE $d/dt \vec{X} = F(\vec{X})$. This equation, known as fluid approximation, can be shown to be a first order approximation of the average of the CTMC, and, under a suitable rescaling of the variables (dividing by the system size, which for biochemical reactions is just the volume), one can prove convergence of the CTMC to the solution of the ODE (see  \cite{tutorial}) as populations and system size go to infinity. Intuitively, this ODE is a good description of the system behaviour when populations are large. 

\paragraph*{Stochastic Hybrid Automata}
In many situations, it is not the case that all entities/ species in the model are present in large quantities. In such scenarios,  fluid approximation can give poor results, yet dealing with CTMCs can be computationally unfeasible. An example are genetic regulatory networks, in which genes are modelled explicitly as a finite state machine \cite{JLC}. In these cases, a better strategy is that of approximating continuously only some variables, keeping discrete the others. This reflects in the dynamics: some reactions will be converted into flows (generally those modifying only continuous variables), while the others will remain stochastic discrete events. This gives rise to a model that can be expressed in terms of a class of Stochastic Hybrid Automata (SHA, \cite{JLC}) known as Piecewise-Deterministic Markov Processes \cite{davis}. Alternatives assuming a stochastic continuous dynamics have also been considered \cite{Ocone:hybrid13,Opper:approximate10}.
More specifically, the SHA so obtained have discrete modes identified by the value of discrete variables. In between discrete transitions, the system evolves following the solution of the differential equation, whose vector field is mode-dependent (via the value of discrete variables). Discrete jumps happen at exponentially random distributed  times, at  a non-constant rate that can depend on the continuous variables. After each jump, the value of discrete variables can change. Also continuous variables can be updated, even if we do not consider this possibility in this paper, see \cite{JLC} for further details. Similarly to the fluid approximation case, we can see SHA models as the limit of CTMC, taking to the limit only the populations corresponding to continuous variables (under a suitable scaling of rates, see \cite{asmta10} for further details).

\paragraph{Signal Temporal Logic.}
Temporal logic~\cite{Pnueli1977} provides a very elegant framework to specify 
in a compact and formal way an emergent behaviour in terms of {\it time-dependent} events.
Among the myriads of temporal logic extensions available, 
 Signal Temporal Logic~\cite{Maler2004} (STL) is very suitable to characterize 
behavioural patterns in time series of real values generated during the simulation of 
a dynamical system. STL extends the dense-time semantics of Metric Interval Temporal 
Logic~\cite{Alur1996} (MITL), with a set of parametrized numerical predicates playing 
the role of atomic propositions. 
STL provides two different semantics: a boolean semantics that returns yes/no 
depending if the observed trace satisfies or not the STL specification and
a quantitative semantics that also returns   measure of robustness of the specification.
Recently, Donz\'e et. al~\cite{Donze2013} proposed a very efficient monitoring algorithm for STL 
robustness, now implemented in the Breach~\cite{Donze2010a} tool. 
The combination of robustness and sensitivity-based analysis of STL formulae
 have been successfully applied in several domains ranging 
from analog circuits~\cite{Jones2010} to systems biology~\cite{Donze2010b,Donze2011}, 
to study the parameter space and also to refine the 
uncertainty of the parameter sets. In the following we recall~\cite{Donze2010} the syntax 
and the quantitative semantics of STL that will be used in the 
rest of the paper. The boolean semantics can be inferred using 
the sign of the quantitative result (positive for true and negative for false).

\begin{defn} [STL syntax] The syntax of the STL is given by
$$\varphi := \top \:|\:  \mu \:|\:  \neg \varphi \:|\: \varphi_{1} \wedge \varphi_{2} \:|\: \varphi_{1} \: \mathcal{U}_{[a,b]} \: \varphi_{2},$$
where $\top$ is a true formula, conjunction and negation are the standard boolean
connectives, $[a,b]$ is a dense-time interval with $a<b$ and  $ \mathcal{U}_{[a,b]}$ is the {\it until} operator. 

The atomic predicate $\mu :  \mathbb{R}^{n} \rightarrow \mathbb{B}$ is defined as $\mu({\bf x}):= (y({\bf x})\geqslant 0)$, where ${\bf x} [t]=(x_{1}[t],...,x_{n}[t])$, $t \in \mathbb{R}_{\geqslant 0}$, $x_{i} \in \mathbb{R}$, is the primary signal,  and $y: \mathbb{R}^{n} \rightarrow \mathbb{R}$ is a real-valued function known as the secondary signal.
\end{defn}

\noindent 
%The time-unbounded operator $\varphi_{1} \: \mathcal{U} \: \varphi_{2}:=\varphi_{1} \: \mathcal{U}_{[0,\infty)} \: \varphi_{2}$  requires $\varphi_{1}$ to hold until $\varphi_{2}$ becomes true.
The (bounded) until operator $\varphi_{1} \: \mathcal{U}_{[a,b]} \: \varphi_{2}$ requires $\varphi_{1}$ 
to hold from now until, in a time between $a$ and $b$ time units, $\varphi_{2}$ becomes true.
The {\it eventually} operator  $F_{[a,b]}$ and  the {\it always } operator $G_{[a,b]}$ can be defined as usual:
 $F_{[a,b]}  \varphi := \top \mathcal{U}_{[a,b)} \varphi$, $G_{[a,b]} \varphi := \neg F_{[a,b]} \neg \varphi.$

\begin{defn} [{\bf STL  Quantitative Semantics} for space robustness]
\label{sp.rob}
\begin{align*}
&\rho(\mu, \vec{x},t)  & = &\mbox{ } \phantom{a}  y(\vec{x}[t]) \qquad \mbox{ where } \mu \equiv y(\vec{x}[t]) \geqslant 0
\\ & \rho (\neg \varphi,\vec{x},t)  & =  & \mbox{ } \phantom{a}  - \rho (\varphi,\vec{x},t)
\\&\rho( \varphi_{1} \wedge  \varphi_{2}, \vec{x},t)  & = &\mbox{ } \phantom{a} \min ( \rho( \varphi_{1},\vec{x},t),\rho( \varphi_{2},\vec{x},t) )
\\& \rho(  \varphi_{1} \: \mathcal{U}_{[a,b)}  \varphi_{2}, \vec{x},t) & = &\mbox{ } \phantom{a}  \max_{t'\in t+[a,b]}(\min(\rho( \varphi_{2},\vec{x},t'),\min_{t'' \in [t,t']}(\rho( \varphi_{1},\vec{x},t''))))
\end{align*}
where $\rho$ is the quantitative satisfaction function, returning a real number $\rho(\varphi, \vec{x},t)$ quantifying the degree of satisfaction of the property $\varphi$ by the signal $\vec{x}$ at time $t$. Moreover, $\rho(\varphi, \vec{x}):=\rho(\varphi, \vec{x},0)$.
\end{defn}

\change{We stress here that the choice of the secondary signals $y:\mathbb{R}^n\rightarrow\mathbb{R}$ is an integral part  of the definition of the STL formula expressing the behaviour of interest. Different choices of secondary signals result in different formulae, hence in different robustness measures.  We also remark that the robustness score of Definition \ref{sp.rob} has to be interpreted as a weight of how much a given model (with fixed initial conditions and parameters) satisfies an STL behaviour. More precisely, its absolute value can be seen as a distance of the signal $\vec{x}$ under consideration from the set of trajectories  satisfying/ dissatisfying  the formula \cite{Fainekos2009}, in STL trajectory are projected with respect to secondary signals $y$ \cite{Donze2010}. In this sense, this measure is different from the more common sensitivity-based notions of robustness, like those discussed  in \cite{komorowski2011}, measuring the size of a region in the parameter space in which the system behaviour is roughly constant. However, sensitivity analysis and its related techniques can  be applied to the robustness score  of Definition \ref{sp.rob}.  }

% !TEX root =  main.tex

\section{Robustness of Stochastic Models}
\label{sec:robustness}

Consider a STL formula $\phi$, with predicates interpreted over state variables of a PCTMC model $\vec{X}(t)$. The boolean semantics of $\phi$ is readily extended to stochastic models as customary, by measuring the probability of the set of trajectories of the CTMC that satisfy the formula:
\[P(\phi) = \bb{P}\{\vec{x}~|~\vec{x}\models \phi  \}. \]

The rationale behind such definition is that a PCTMC model defines a probability distribution on the space of trajectories, which is usually obtained by applying the cylindric construction \cite{Baier2003}. Furthermore, the set of trajectories that satisfy/ falsify a formula is a measurable set, so that we can safely talk about its probability. In the following, we will refer to the space of trajectories as $\cal{D}$, and interpret the PCTMC model $\vec{X}(t)$ as a random variable $\vec{X}$ over $\mathcal{D}$.
In order to extend this definition to the robustness score, it is convenient to think of the set of trajectories that satisfy $\phi$ as a measurable function $I_\phi:\cal{D}\rightarrow \{0,1\}$, such that $I_\phi(\vec{x})  =1$ if and only if $\vec{x}\models \phi$. Then, we can define the random variable $I_\phi(\vec{X})$ on $\{0,1\}$ induced by the PCTMC $\vec{X}$ via $I_\phi$ as the Bernoulli random variable which is equal to 1 with probability $P(\phi)$. We can equivalently write: 

\[\bb{P}(I_\phi(\vec{X}) = 1) = \bb{P}(\{\vec{x}\in\cal{D}~|~  I_\phi(\vec{x})=1\} ) = \bb{P}(I_\phi^{-1}(1))\]

We can extend the robustness score to PCTMC models in the same way: given a trajectory $\vec{x}(t)$, we can compute its robustness score according to Def. \ref{sp.rob} and interpret $\rho(\phi,\vec{x},0)$  as a function from the trajectories in $\cal{D}$ to $\bb{R}$. This function is easily seen to be measurable \change{(with respect to the $\sigma$-algebra induced from the Skorokhod topology in $\mathcal{D}$)}, and so it induces a real-valued random variable $R_{\phi}(\vec{X})$ with probability distribution given by
 \[\bb{P}\left(R_{\phi}(\vec{X})\in [a,b] \right)= \bb{P}\left( \vec{X} \in \{\vec{x}\in\cal{D}~|~\rho(\phi,\vec{x},0)\in [a,b] \}   \right)  \]

%
%In order to extend this definition to the robustness score, we will more generally consider probability distributions on the function space of cadlag functions. 
%A cadlag function from $[0,T]$ in $\bb{R}^n$ is a function with values in $\bb{R}^n$ which is right continuous and has left limits for any point $t\in[0,T]$.  Note that we consider bounded time horizons because STL properties are time bounded. We call $\cal{D}_{[0,T]}(\bb{R}^n)$  the  space of cadlag functions, which turns to be a complete metric space under the Skorohod metric \cite{BILLINGSLEY}.
%
%

%
%Consider a cadlag function $\vec{f}\in \cal{D}_{[0,T]}(\bb{R}^n)$ and an STL formula $\phi$. Then we can interpret $\phi$ on the trajectory $\vec{f}(t)$, and compute the robustness score $\rho(\phi,\vec{f},0)\in\bb{R}$. Hence, we can interpret $\rho_\phi(\vec{f}) =  \rho(\phi,\vec{f},0)$ as a \emph{functional} from $\cal{D}_{[0,T]}(\bb{R}^n)$  to $\bb{R}$. If instead of considering a single trajectory we consider a probability distribution on the trajectory space, as is the case for PCTMC models, we can lift the robustness function to probability measures:  $\rho_\phi(\cdot)$ turns a distribution on trajectories into a distribution on real numbers, i.e. it assigns to each random variable  $\vec{X}$ on $\cal{D}_{[0,T]}(\bb{R}^n)$  a real-valued random variable $R_\vec{X} = \rho_\phi(\vec{X})$. In particular, we have that \[\bb{P}\left(R_\vec{X}\in [a,b] \right)= \bb{P}\left( \vec{X} \in \{\vec{f}~|~\rho_\phi(\vec{f})\in [a,b] \}   \right) \]

Staten otherwise, if we apply the definition of robustness to a stochastic model, we obtain a distribution of robustness degrees. This distribution tells us much more than the standard probabilistic semantics, because it tells us ``how much'' a formula is true. 

In particular, in this paper we will be interested in some statistics of this distribution, specifically the average robustness degree, and the average robustness conditional on a formula being true or false. The first quantity gives a measure of how strongly a formula is satisfied on average. The larger this number, the more robust is satisfaction. Most of the times, this number will be correlated with the satisfaction probability, yet we can have a large average satisfaction score even for a small probability of satisfaction. Better indicators of the intensity of satisfaction and dissatisfaction are the conditional averages, $\bb{E}(R_{\phi}~|~R_{\phi}>0)$ and $\bb{E}(R_{\phi}~|~R_{\phi}<0)$. These are related to the average by the equation
\[\bb{E}(R_{\phi}) = P(\phi) \bb{E}(R_{\phi}~|~R_{\phi}>0) + (1-P(\phi)) \bb{E}(R_{\phi}~|~R_{\phi}<0) \]
which holds provided $\bb{P}(R_{\phi} = 0)$ is zero.
 
One goal of this paper is to investigate to what extent these three synthetic indices are good descriptors of the robustness distribution, and how they can be exploited to do parameter synthesis for PCTMC models.

% !TEX root =  main.tex

\begin{table}[!t]
\begin{center}
\begin{tabular}{|l|l|l|}
\hline
Reaction & rate constant & init pop\\
\hline
$A + 2 X \rightarrow 3 X$ & $k_1 = 3 \cdot10^{-7}$ & $X(0) = 247$ \\
\hline
$3 X \rightarrow A + 2 X$ & $k_2 = 1 \cdot10^{-4}$ & $A(0) = 10^5$\\
\hline
$B \rightarrow X$ & $k_3 = 1 \cdot 10^{-3}$ & $B(0) = 2\cdot 10^5$\\
\hline
$X \rightarrow B$ & $k_4 = 3.5$ & \\
\hline
\end{tabular}
\end{center}
\caption{Biochemical reactions of the Schl\"ogl model. Parameters are taken from \cite{sensitivityCTMC}.}
\label{tab:schlogl}
\end{table}

\section{Case Studies}
\label{sec:casestudy}

%
%
%The idea is to study two models:
%a model of CTMCs, of a bistable system. I have one in mind (shred, or something like it, I will find it - it is a chemical system). The nice thing is that we can show that this analysis of robustness captures bistability: the robustness score distribution will be bistable.
%
%The other system has to be a stochastic hybrid system. I have in mind one model of the circadian clock, for which a stochastic hybrid model oscillates. Then we can study robustness of oscillations!

In this section, we investigate experimentally the notion of robust semantics of STL formulae for stochastic models. We will consider two systems: the Schl\"ogl system \cite{petzold05}, a simple network of biochemical reactions exhibiting a bistable behaviour, and the Repressilator \cite{elowitz00}, a synthetic biological clock implemented as a network of gene regulations. More specifically, we consider a CTMC model of the Schl\"og system and a hybrid model of  the Repressilator \cite{lucaIFAC07,lucaTCS10}, in order to illustrate the general applicability of the stochastic robust semantics introduced in Section \ref{sec:robustness}.

\begin{figure}[!t]
\begin{center}
\subfigure[CTMC simulation]{\label{fig:schloglSim}
\includegraphics[width=.43\textwidth]{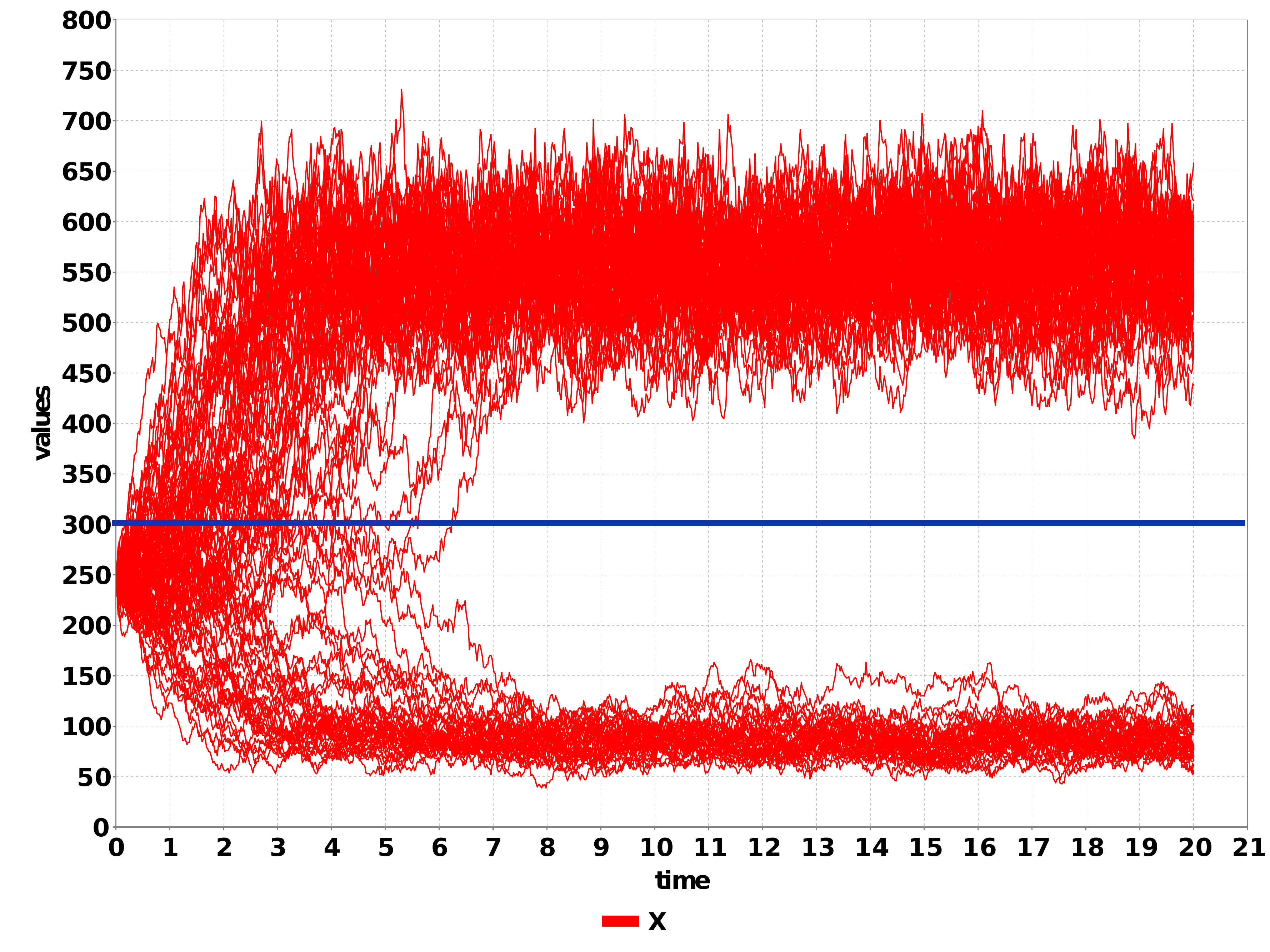} }
\vspace{5cm}
\subfigure[robustness distribution]{\label{fig:bistable}
\includegraphics[width=.50\textwidth]{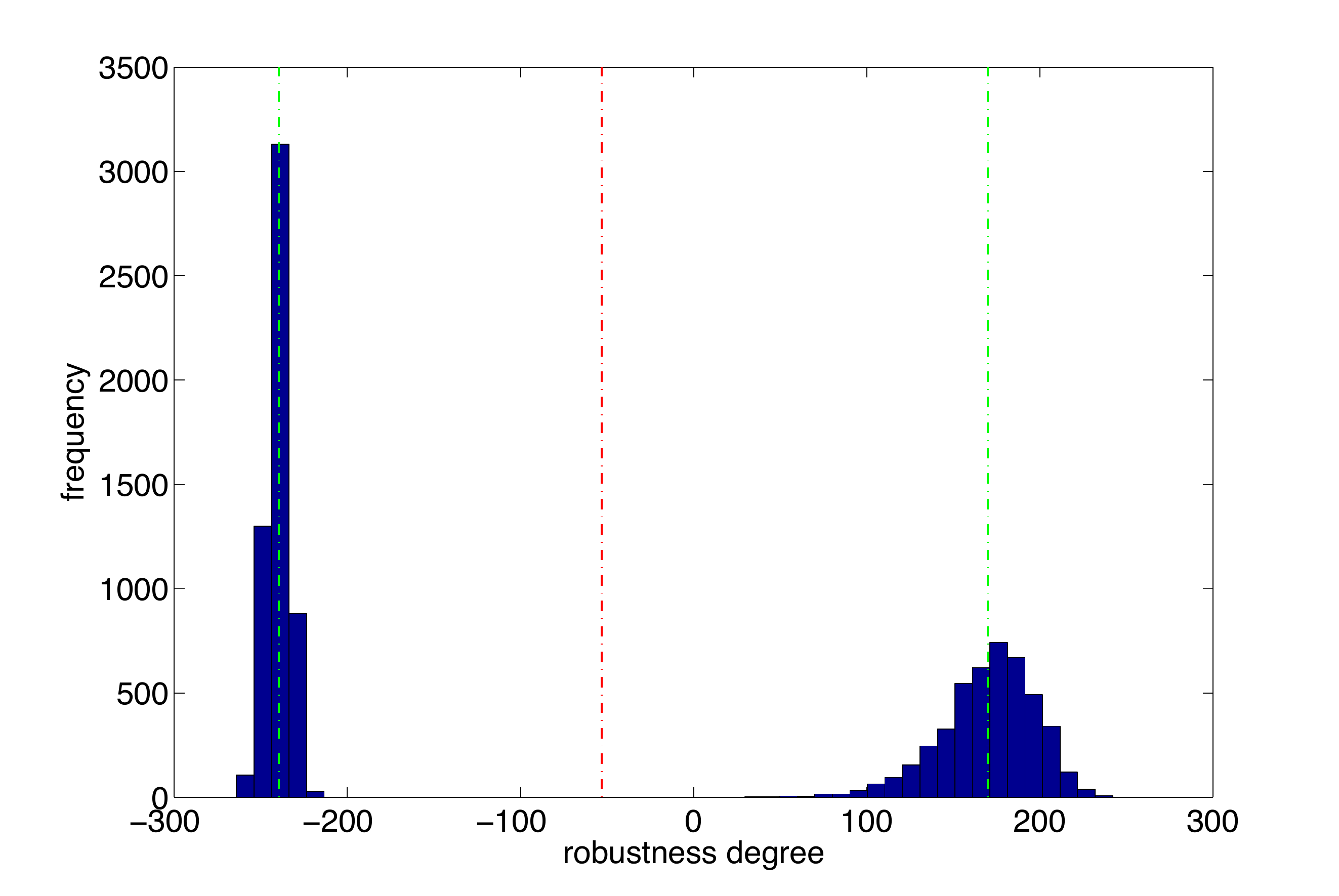} }
\end{center}
\vspace{-5cm}
\caption{Simulation of the Schlogl model (100 runs), for parameters as in Table \ref{tab:schlogl}. The blue straight line is the value $X=300$ (left). The robustness distribution of the STL formula \ref{STL:bistable} with $T_{1}=10$ and $T_{2}=15$ time units. It has  average robustness  -53.15  (vertical red line), conditional robustness 169.89 and -239.52 (vertical green lines), and satisfaction probability  0.4552 (right).}
\end{figure}

\subsection{Schl\"ogl system}
\label{sec:schlogl}
The Schl\"ogl model is a simple biochemical network with four reactions, listed in Table \ref{tab:schlogl}. The rates of the reactions are computed according to the mass action principle for stochastic models \cite{gillespie}. Species $A$ and $B$ are considered to be present in large quantities, hence assumed constant. The characteristic of this system is to have, for certain parameter values, like the one shown in Table \ref{tab:schlogl}, a bistable behaviour. More specifically, the reaction rate ODE system has two stable steady states, and for this model the trajectories of the stochastic system starting from a fixed initial state $x_0$  can end up in  one attractor or the other. The probability of choosing one stable state or the other  depends on the position of $x_0$ relatively to the basin of attraction of the two equilibria. If we start close to its boundary, the bistable behaviour becomes evident, see Figure \ref{fig:schloglSim}. 

We now consider the property of eventually ending up in one basin of attraction, and express it with the STL formula
\begin{equation}
\label{STL:bistable} 
\varphi: \mathbf{F}^{[0,T_1]}\mathbf{G}^{[0,T_2]} (X \geq k_{t})\ \ \ k_{t} = 300 
\end{equation}
stating that the system, after at most $T_1$ units of time, stabilises to a value which remains above $k_{t}=300$ for as long as $T_2$ units of time.  \change{In this formula, the predicate $\mu(X) = X \geq k_{t}$ corresponds to the linear secondary signal $y(X) = X-k_t$.}  As can be seen from Figure \ref{fig:schloglSim}, if the model is in the large equilibrium, then this property will be true, and false in the other case.

If we estimate the probability of the formula statistically, then we obtain the value $p=0.4583$ (10000 runs, error $\pm 0.02$ at $95\%$ confidence level). However, this raw number does not tell us anything specific about bistability. A system stabilising just above the threshold 300, such that roughly $55\%$ of its trajectories cross it ``frequently'', may satisfy the same formula with the same probability. 
However, the bimodal behaviour becomes evident if we look at the distribution of the robustness degree of the formula, see  Figure \ref{fig:bistable} . Hence, the robustness score carries an additional amount of information with respect to \change{the satisfaction probability of a STL formula. We stress  that we are not comparing the robustness degree with the probability distribution of the CTMC $X[t]$: both the satisfaction probability of $\varphi$ and its robustness are (unidimensional) quantities derived from $X[t]$, which are easier to compute and visualise. } 

\begin{figure}
\begin{center}
\subfigure[varying threshold]{\label{fig:threshold}
\includegraphics[width=.48\textwidth]{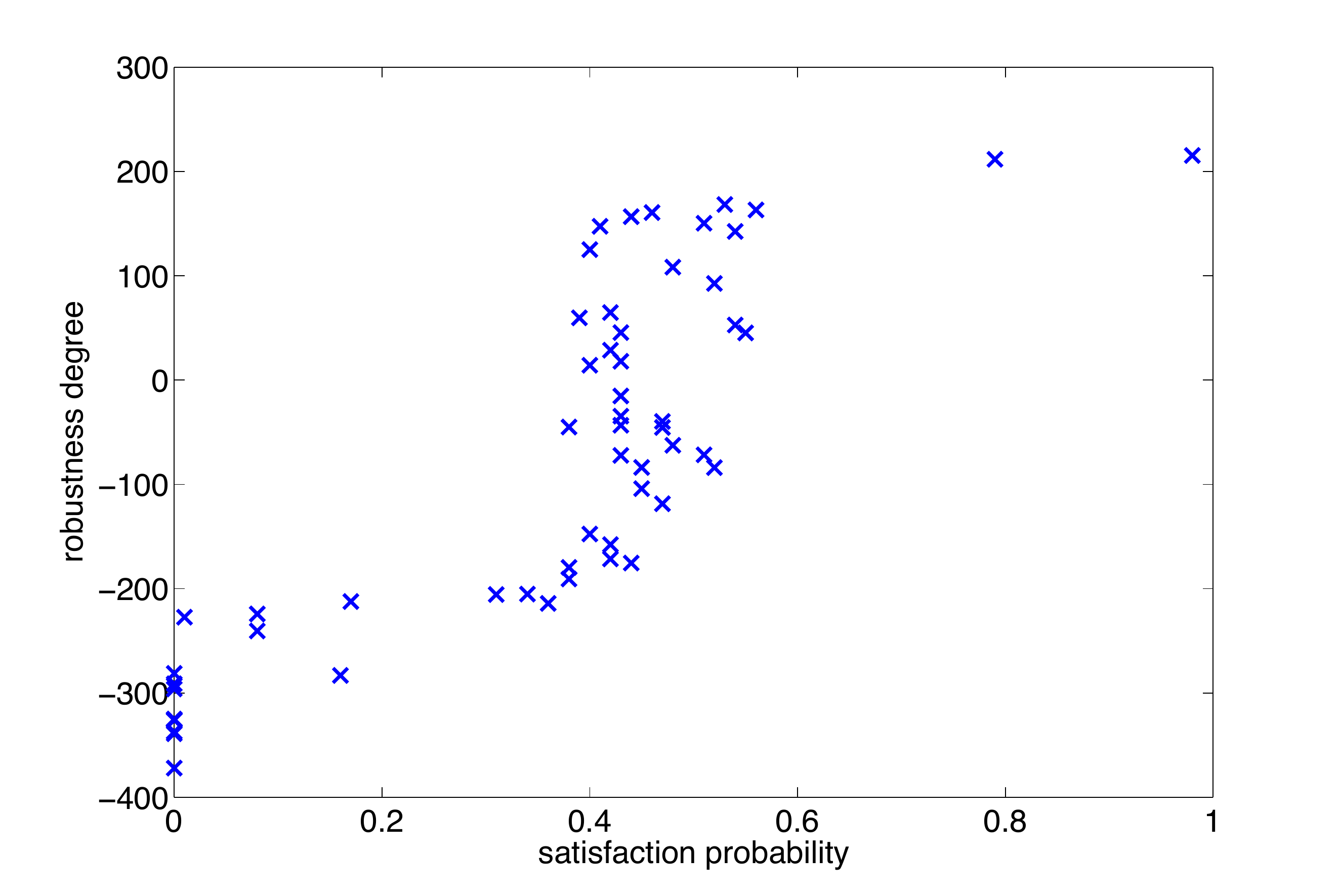} }
\subfigure[varying $k_3$]{\label{fig:k1}
\includegraphics[width=.48\textwidth]{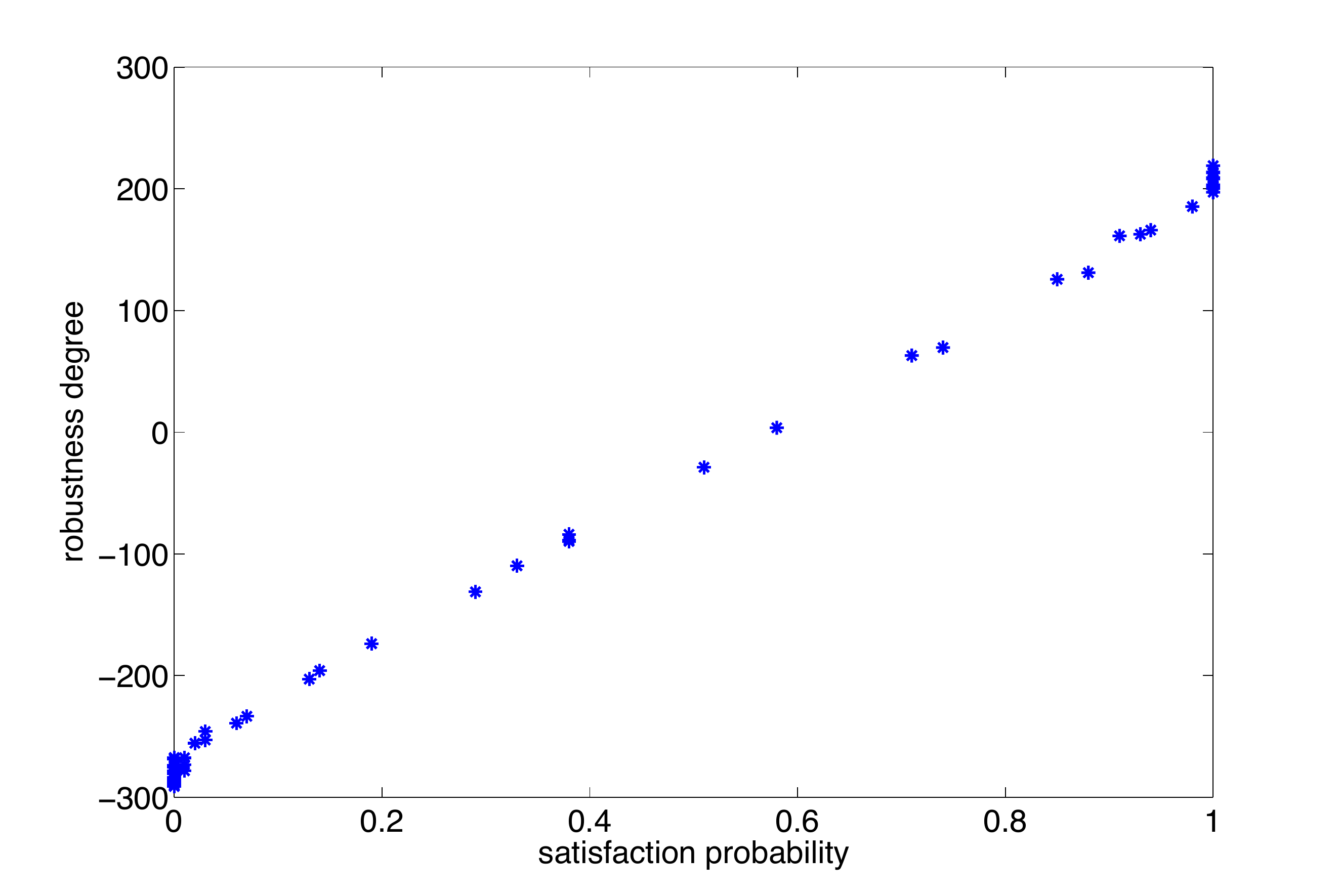} }
\end{center}
\caption{Satisfaction probability versus average robustness degree for varying (left) threshold $k_{t}$ in the STL formula (\ref{STL:bistable}) and (right) parameter $k_3$. $k_3$ was varied between 100 and 300 in steps of $10$ units, while the threshold was varied between 50 and 600 in steps of 10. }
\label{fig:averageVSprob}
\end{figure}

In Figure \ref{fig:averageVSprob}, we investigate the behaviour of the average robustness degree, and its relationship with the satisfaction probability. In order to do this, we varied the threshold level $k_{t}$ in the formula (Figure \ref{fig:threshold}), and the rate constant $k_3$ (Figure \ref{fig:k1}), \change{and estimated statistically the average robustness degree and the satisfaction probability from 10000 runs for each parameter combination.}  As we can see these two quantities are correlated. 
When we vary the threshold, the correlation between satisfaction probability and robustness score is around 0.8386, while the dependency seems to follow a sigmoid shaped curve.  In the second case, instead, the correlation between satisfaction probability and average robustness degree is 0.9718, with an evident linear trend.

Finally, we consider the conditional robustness degrees. For model parameters as in Table \ref{tab:schlogl}, the average robustness conditional on  Formula (\ref{STL:bistable}) being true is 169.89, while the robustness conditional on the formula being false is -239.52 (see also Figure \ref{fig:bistable}). These two indicators estimate how robustly the system remains in the basin of attraction of each steady state.

\subsection{Repressilator}
\label{sec:repressilator} 

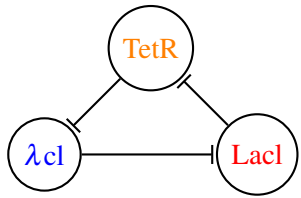
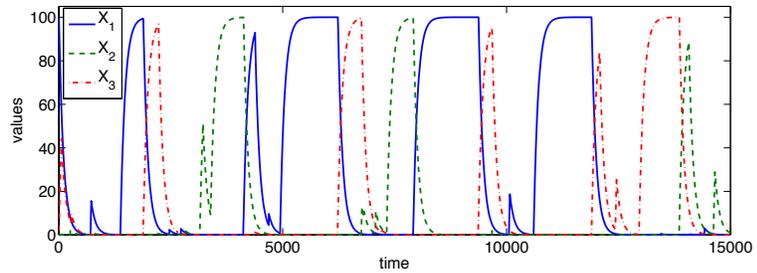
\begin{figure}[!th]
\begin{center}

\subfigure[Repressilator, gene network]{ \label{fig:Rep} 
\begin{tikzpicture}[->,>=stealth',shorten >=1pt,auto,node distance=2.cm,on grid,thick]
  \node[state,fill = white!50!white]  (A)                    {{\color{orange}TetR}};
  \node[state,fill = white!50!white] (B) [below left =of A]    {{\color{blue}$\lambda$cl}};
    \node[state,fill = white!50!white] (C) [below right =of A]    {{\color{red}Lacl}};
%\tikzset{mystyle/.style={->,double=orange}}
%\tikzset{every node/.style={fill=white}}
\path (A) edge [-|,right] node {$$} (B)
      (B) edge [-|,below] node {$$} (C)
      (C) edge [-|,left] node {$$} (A);
\end{tikzpicture}
 }
\subfigure[Hybrid stochastic simulation]{\label{fig:RepSim}
\includegraphics[width=.72\textwidth]{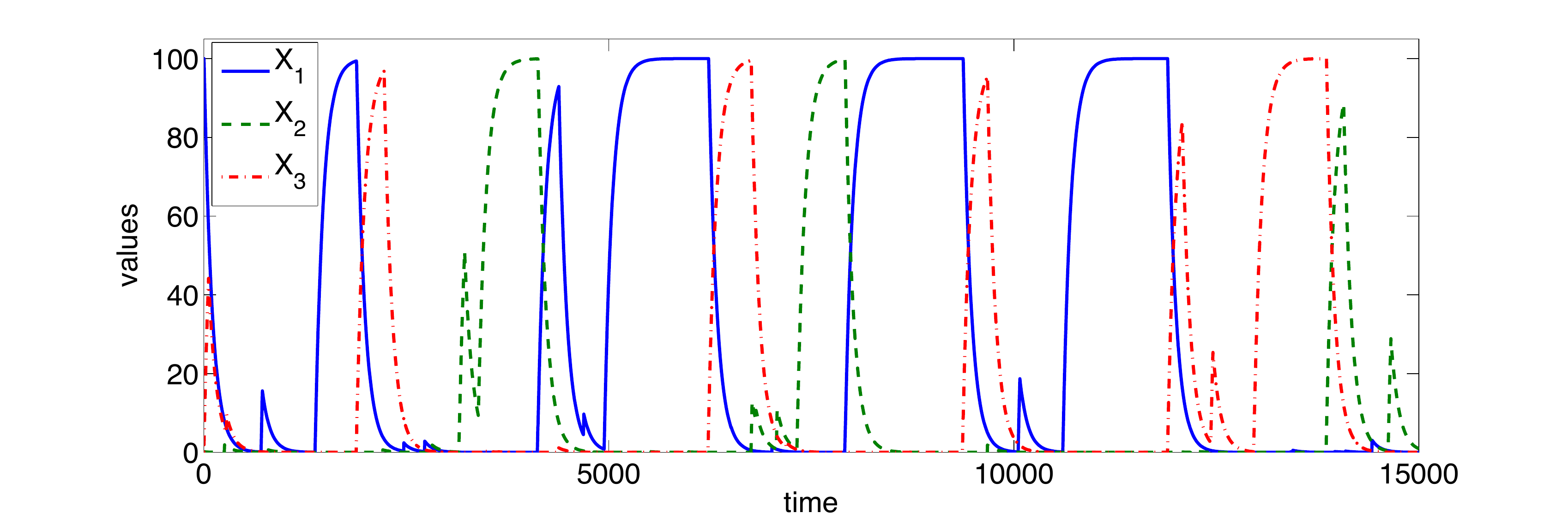} }
\end{center}

\caption{The repressilator (left) is a cyclic negative-feedback loop composed of three repressor genes: TetR, $\lambda$cl, Lacl.  Oscillatory behaviour of the model (right), for model parameters: protein production rate $k_p = 1$, protein degradation rate $k_d = 0.01$, repressor binding rate $k_b = 0.1$, repressor unbinding rate $k_u = 0.001$.   }
\label{fig:Rep}
\end{figure}

The second case study is a genuine stochastic hybrid model of the Repressilator \cite{elowitz00}, a synthetic genetic clock composed of three genes expressing three transcription factors repressing each other in a cyclical fashion (see Figure \ref{fig:Rep}). The stochastic hybrid model we consider is taken from \cite{lucaIFAC07,lucaTCS10}. In the model, we lump the transcription and translation in a single event, and model production and degradation of the protein as continuous flows. The binding and unbinding of transcription factors from gene promoters, instead, are modelled as discrete and stochastic events. As we can see in Figure \ref{fig:RepSim},  the model exhibit sustained oscillations, albeit with an irregular period. This happens for parameters giving a strong repression via a low unbinding rate.

In order to check for the presence of oscillations, we use the STL formula
{\small 
\begin{equation}
\mathbf{G}^{[0,T]}( ((X_i < k_{low}  ) \rightarrow \mathbf{F}^{[T_1,T_2]}(X_i > k_{high}  ) )\wedge ((X_i > \rho_{high}  ) \rightarrow \mathbf{F}^{[T_1,T_2]}(X_i < k_{low}  ) ) \wedge \mathbf{F}^{[0,T_2]}(X_i > k_{high}  )  ), 
\label{prop repress}
\end{equation}}
expressing the fact that low values of $X_i$ alternate to high values, with a period between $T_1$ and $T_2$. \change{The secondary signals are $k_{low} - X_i$, $X_i-k_{high}$, and so on.} Here $X_i$ can be one of the three proteins of the Repressilator. In the next discussion, we focus on $X_1$. 

\begin{figure}[!th]
\begin{center}
\subfigure[Robustness distribution ]{\label{fig:RepDist}
\includegraphics[width=.485\textwidth]{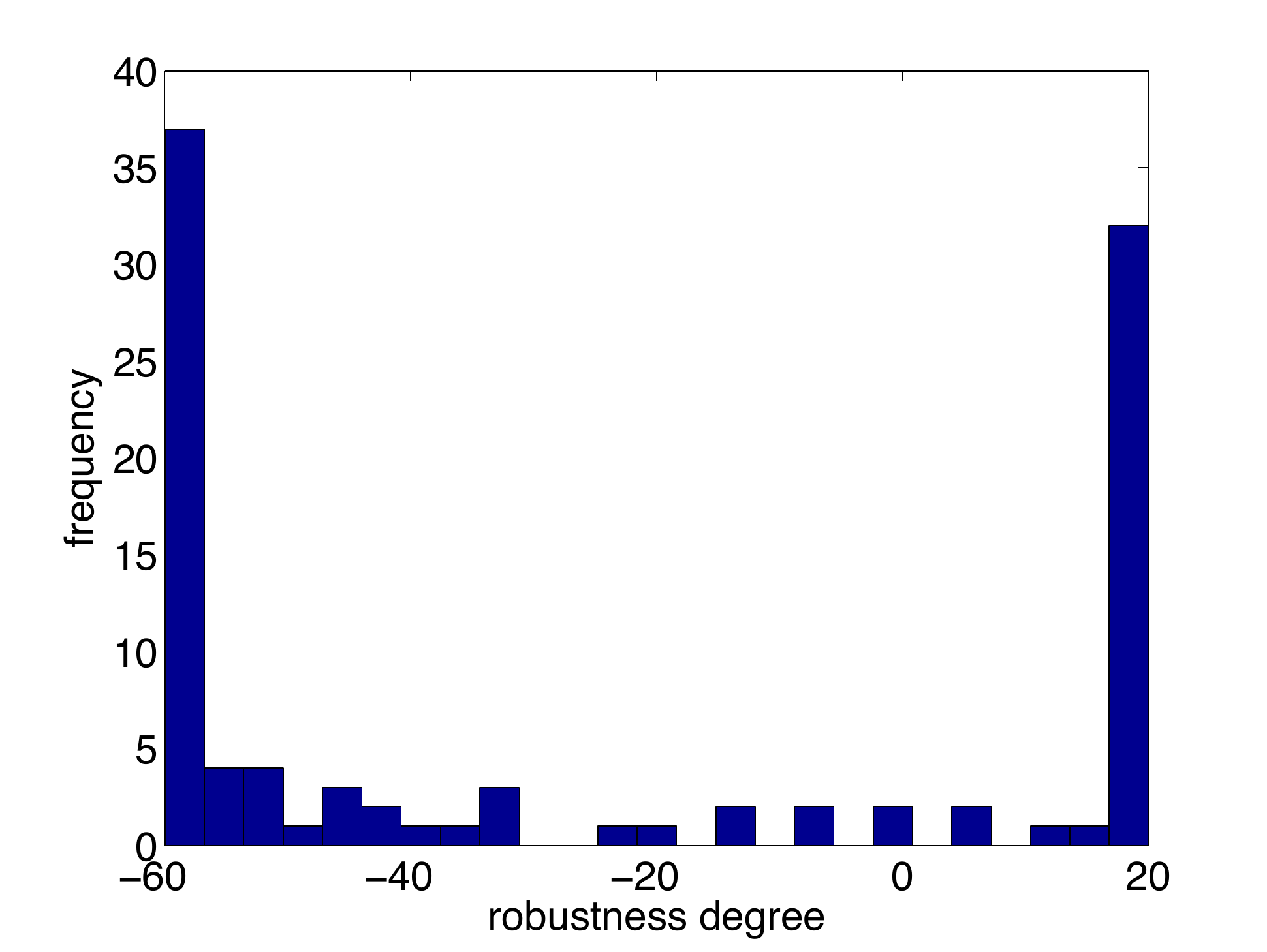} }
\subfigure[satisfaction probability vs robustness degree]{\label{fig:thresholdpropT2}
\includegraphics[width=.48\textwidth]{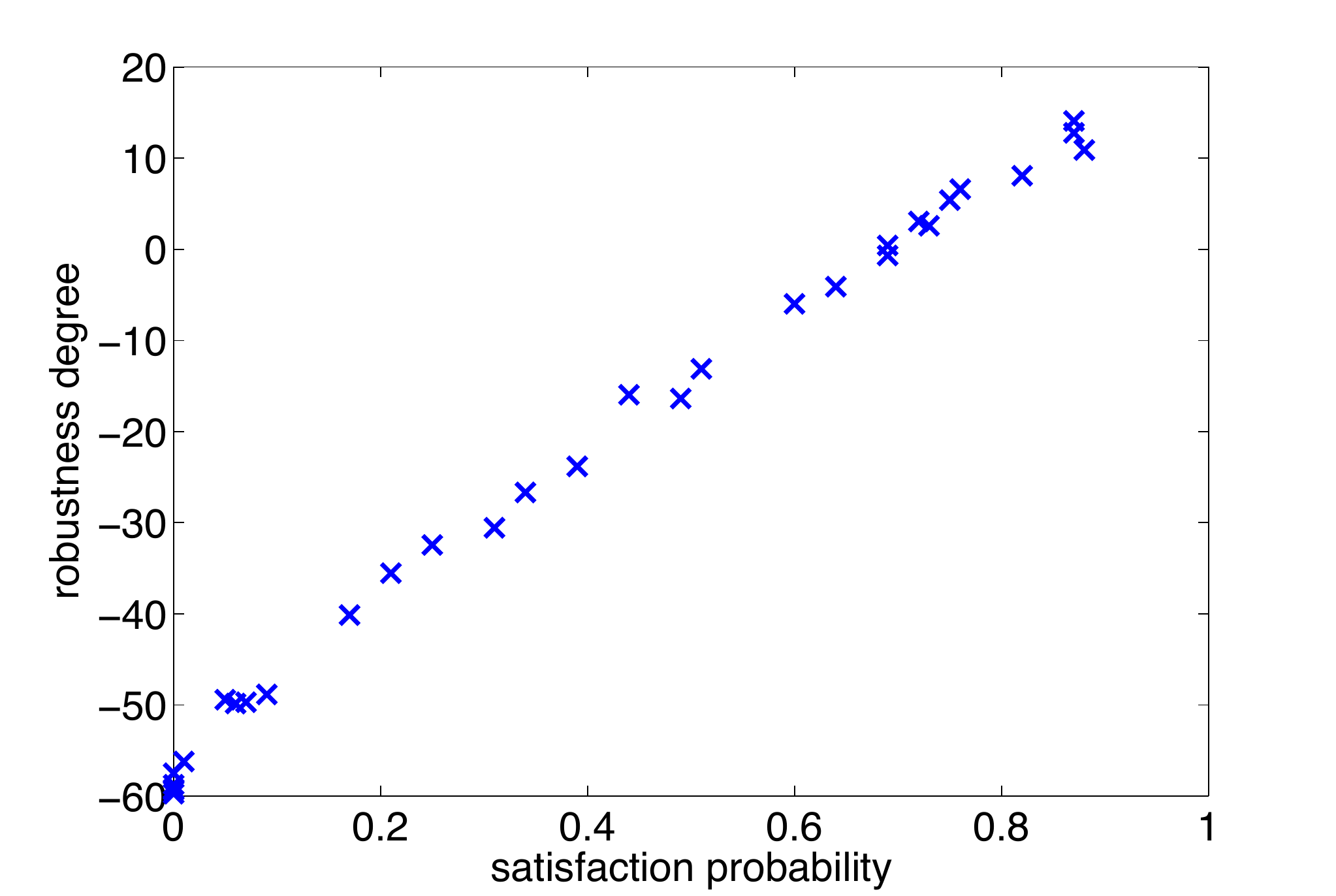} }
\end{center}
\caption{ Robustness distribution for Formula \ref{prop repress} parameters $k_{low}=20$, $k_{high}=60$, $T_{1}=100$, $T_{2}=4000$, $T=7000$. Average robustness is  -25.888 and estimated satisfaction probability  is 0.35 (left). Satisfaction probability versus average robustness degree. $T_{2}$ was varied between 1000 and 7000 in steps of 100 units (right). }
\vspace{-0.4cm}

\end{figure}

Again, the robustness score gives us a measure of the satisfaction/dissatisfaction of the formula.
As we can see from Figure \ref{fig:RepDist}, the robustness degree shows a bimodal behaviour also in this case. 
 In particular, in case the formula is false, it gives some degree of information on the amplitude of oscillations, and on the stability of the period \change{(relatively to the formula parameters)}. In fact, a robustness value of, say, -50 can be obtained for instance if from a point in which $X_i < k_{low}$, the system remains below $k_{high}-50$ for a whole (half) period of the oscillation (which is constrained to be in $[T_1,T_2]$). This can happen due to low amplitude or irregular period. 
In Figure \ref{fig:thresholdpropT2}, we plot the average robustness degree against the satisfaction probability, varying the  property parameter $T_{2}$, showing once again the correlation between the two quantities.

%
%\begin{figure}
%\begin{center}
%\subfigure[T2 versus robustness degree]{\label{fig:thresholdT2}
%\includegraphics[width=.48\textwidth]{figs/repress_robustness_T2.pdf} }
%
%
%
%\end{center}
%\caption{Varying $T_{2}$, with $T_{1}=0$,:   (left), parameter $T_{2}$ versus robustness degree (left) ;  }
%\label{fig:averageVSprob}
%\end{figure}

% !TEX root =  main.tex

\section{System Design}
\label{sec:design}

%**SKETCHED**

We now discuss an application of the robust semantics to the system design problem. The problem we want to tackle is the following:

\noindent \emph{given a population (hybrid) model, depending on a set of parameters $\boldsymbol{\theta}\in \boldsymbol{K}$, and  a specification $\phi$ given by a STL formula, find the parameter combination $\boldsymbol{\theta^*}$ such that system satisfied $\phi$ with probability at least $p\in[0,1]$ as \emph{robustly} as possible.}
\\
We will tackle this problem by:
\begin{itemize}
\item rephrasing it as an unconstrained  optimisation problem, where we seek to optimise the average robustness, using penalty terms to encode for probability constraints. More specifically, assuming we want to enforce the satisfaction probability to be at least $q$, we add a penalty term of the form $\alpha \|p-q\|$, if $p < q$, and 0 otherwise, where $\alpha<0$ controls the penalty intensity.  
\item evaluating the function to optimise using statistical model checking with a fixed number of runs, usually set to 100;
\item solving the optimisation problem using an optimisation strategy for reinforcement learning, based on statistical emulation and Gaussian processes regression (Gaussian Process - Upper Confidence Bound optimisation, GP-UCB \cite{gpucb}).
\end{itemize}
%
%In particular, the choice of the optimisation strategy allows us to solve the problem with a small number of function evaluations, taking also into account the noise in the computed value of functions, due to statistical estimation. Note that GP-UCB has been used before in the context of statistical model checking in \cite{lucaQEST13}, as a routine to find maximum likelihood estimated of model parameters given qualitative observations.
%In the following, we will introduce briefly the optimisation algorithm, then presenting its application to the two case studies of the paper.

\subsection{Gaussian Processes - Upper Confidence Bound Optimisation}
\label{gpucb}

% !TEX root =  hsb_2013.tex

\paragraph*{Gaussian Processes.}

The key ingredient for the design problem is an efficient estimation of the unknown objective function, i.e. the average robustness as a function of the process kinetic parameters. Function approximation is a central task in machine learning and statistics. The general regression task can be formulated as follows~\cite{Bishop2006}: 
%\cite[e.g.][]{Bishop2006}: 
given a set of input-output pairs $(\mathbf{x}_i,y_i), \quad i=1,..,N$ ({\it training data}), with $\mathbf{x}_i\in\mathbb{R}^d$ and $y_i\in\mathbb{R}$, determine a function $f\colon\mathbb{R}^d\rightarrow\mathbb{R}$ s.t. $f(\mathbf{x}_i)$ is optimally close to the target values $y_i$ (usually in terms of minimising a suitable loss function). Several methods exist for addressing this task; in this paper we consider Gaussian Process (GP) regression, a popular Bayesian methodology \cite{Rasmussen2007}.
%\cite[e.g.][]{Rasmussen2007}. 
GPs are flexible non-parametric distributions over spaces of functions which can be used as prior distributions in a Bayesian framework, where the input-output pairs represent noisy observations of the unknown function. This enables a natural quantification of the uncertainty of the estimated function at every new input value; this uncertainty will play a central role in the optimal design strategy we propose in Section \ref{sec:design}.
We now give a semi-formal definition of GP \cite{Rasmussen2007}:
\begin{definition} A Gaussian Process over a (portion of) $\mathbb{R}^d$ is a collection of random variables indexed by $\mathbf{x}\in\mathbb{R}^d$ such that every finite dimensional marginal distribution is multivariate normal. Furthermore, there exist two functions $\mu\colon\mathbb{R}^d\rightarrow\mathbb{R}$ ({\it mean function}) and $K\colon\mathbb{R}^d\times\mathbb{R}^d\rightarrow\mathbb{R}$ (covariance function) such that the mean and covariance of the finite dimensional normal marginals is given by evaluating the mean and covariance functions at each point and each pair of points respectively.\end{definition}
We denote a sample from a GP with mean function $\mu$ and covariance function $K$ as\[
f\sim\mathcal{GP}(\mu,K).\]

In practice, the input-output pairs in a regression task are often different features of  experimentally observed data points. In this paper, the output points correspond to true functional evaluations of an unknown (and analytically intractable) function of the inputs. In this case, the regression task is often given the special name of {\it emulation} in the statistics literature: the true (but unknown) function is assumed to be a draw from a GP, and the functional evaluations are used as observations to obtain a posterior estimate of the unknown function. This approach was initially introduced in order to perform sensitivity analysis for deterministic computer models in \cite{Kennedy2001}; in that case, the function evaluations could be assumed to be noiseless (apart from numerical errors that were considered negligible in that paper). In our case, the function linking model parameters to average robustness cannot be computed, and we can only obtain a sampling approximation through a Statistical Model Checking procedure. This means that our function evaluations will be noisy; by virtue of the Central limit theorem we can assume that, provided sufficient samples were used for the SMC estimates, the noise in the observed robustness estimates will be approximately Gaussian.\footnote{\change{Note that here we approximate as a GP the average robustness score (or any other fitness score) as a function of parameters. We are not imposing any (Gaussian) approximation of the process itself.}} This therefore enables us to obtain an analytical estimate of the posterior process \cite{Rasmussen2007}. Furthermore, the SMC samples also allow us to estimate the (sample) variance in the average robustness at every sampled parameter value; this information can also be included leading us to a heteroscedastic (i.e. with non identical noise) regression problem (which is however still analytically tractable).

% !TEX root =  hsb_2013.tex

\renewcommand{\Theta}{\boldsymbol{\theta}}
\paragraph*{Gaussian Processes Optimization: the GP-UCB algorithm}

As we have seen, GP emulation provides a convenient way to explore approximately the average robustness of a stochastic process for different values of the model parameters. One could then be tempted to also use the emulated robustness profile for model design, i.e. find the optimum of the emulated function. This strategy, while appealing in its simplicity, is vulnerable to local optima: the emulated function is estimated based on relatively few function evaluation, so that, while the emulator typically provides a good approximation of the true function near the sampled points, regions of parameter space far from the sampled points may contain the true maximum undetected. Using the language of reinforcement learning, maximising the emulated function would privilege exploitation (i.e. using currently available information) at the expense of exploration. Obviously, given sufficient computational power, one may consider sampling many parameter points so as to have sufficient coverage of the whole region of interest; this strategy is however bound to fail in even moderate dimensions due to the curse of dimensionality.

An elegant solution to the above conundrum can be obtained by also considering the {\it uncertainty} of the emulated function (which is also computed analytically in GP regression): intuitively, one should explore regions where the maximum {\it could plausibly be}, i.e. regions in parameter space where there is substantial posterior probability mass for the function to take a high value. We formalise these ideas in a recursive search rule, the so called Gaussian Process Upper Confidence Bound (GP-UCB) algorithm: assume we have computed the average robustness at $N$ parameter values (so that we have $N$ input output pairs). Let $\mu_N(\Theta)$ and $\nu_N(\Theta)$ be the mean and variance of the GP emulator at a given point in input space $\Theta$ (recall that the marginal at any point will be Gaussian)\footnote{We now denote the input as $\Theta$ to emphasize that they are the parameters of a stochastic process}. We select the parameter value $\Theta_{N+1}$ for the next function evaluation according to the following rule\begin{equation}
\Theta_{N+1}=\mathrm{arg max}_{\Theta}\left(\mu_N(\Theta)+\beta_{N+1}\nu_N(\Theta)\right)\label{GPUCBrule}\end{equation}
where $\beta_{N+1}$ is a parameter. Thus, the next point for exploration does not maximise the emulated function, but an upper confidence bound at a certain confidence level specified by the parameter $\beta_{N+1}$ (the quantile can be obtained by applying the inverse probit transform to the parameter).
\cite{gpucb} proved that this algorithm converges to the global maximum of the unknown function with high probability (which can be adjusted by varying the algorithm's parameters). 
%More specifically, they proved the following
%%%%THIS THEOREM CAN BE SAFELY REMOVED IF SPACE IS NEEDED
%\begin{thm}
%Let $\beta_t=k+\alpha\log t$, where $k$ and $\alpha$ are positive constants. Denote as $f$ the unknown function and with $x^*$ the argument where it attains its global maximum. Then, with high probability, the GP-UCB rule of equation \eqref{GPUCBrule} will converge and the following property holds ({\it no regret})\[
%\lim_{T\rightarrow\infty}\frac{1}{T}\sum_{t=1}^T\left(f(x^*)-f(x_t)\right)=0.\]
%\end{thm}

%The GP-UCB algorithm provides a provably convergent approach for global optimisation. 
The primary difficulty in applying GP-UCB is that, in order to be able to apply the rule, the emulated function must be computed at a large number of points; while this is obviously not as onerous as evaluating the true unknown function (as the emulator is known analytically), it may still be problematic for high dimensional parameter spaces. Nevertheless, the algorithm can be applied effectively for moderate sized parameter spaces (of the order of 10 parameters), and modular construction may be used to extend to higher dimensional systems \cite{anastasisHSB12,Bartocci2013}.

\subsection{Experimental Results}

\begin{figure}
\begin{center}
\subfigure[The estimated robustness function]{\label{fig:robustnessFunctionSchlog}
\includegraphics[width=.498\textwidth]{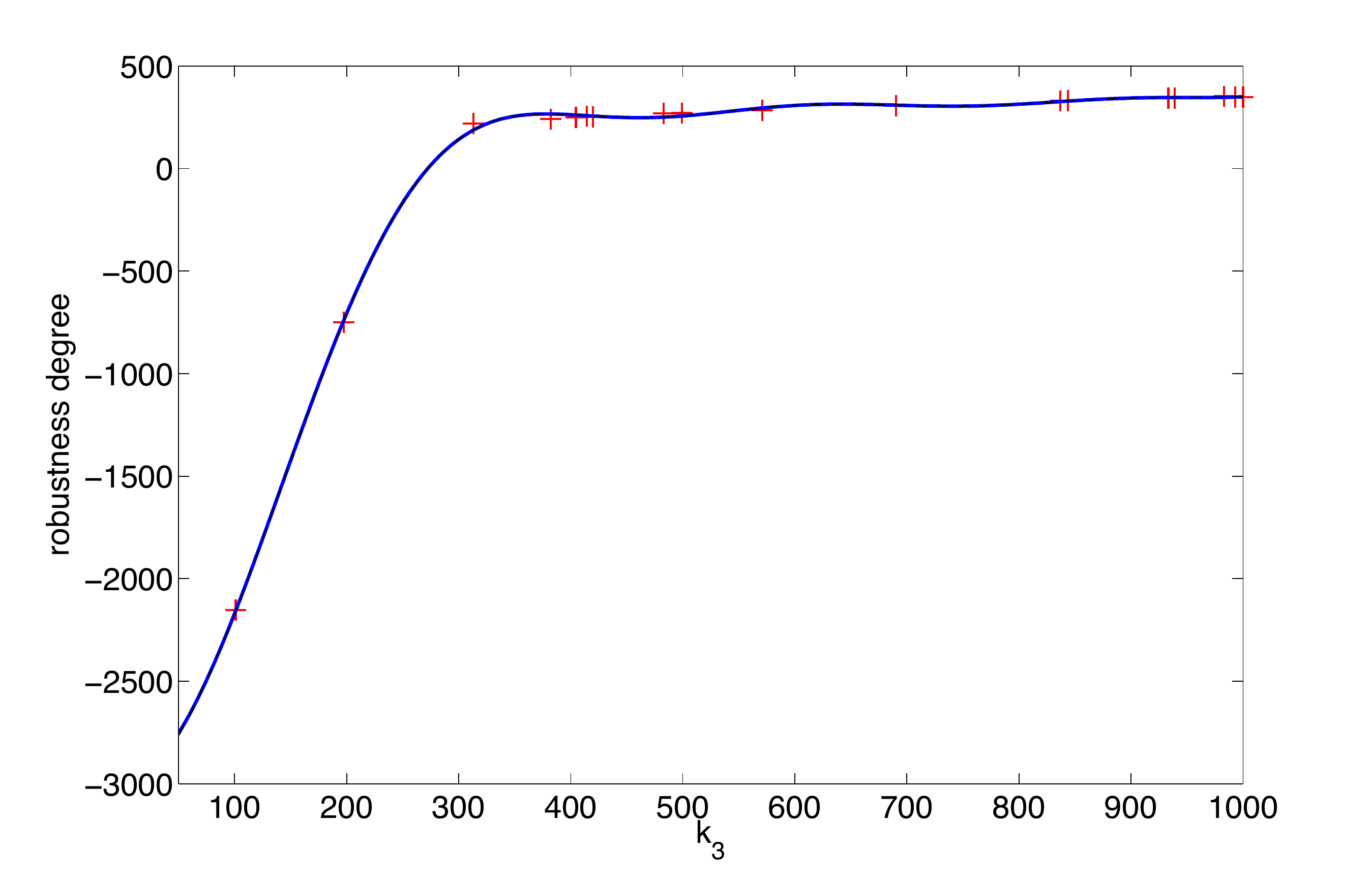} }
\subfigure[Robustness distribution after the optimization]{\label{fig:robustnessSchlog}
\includegraphics[width=.472\textwidth]{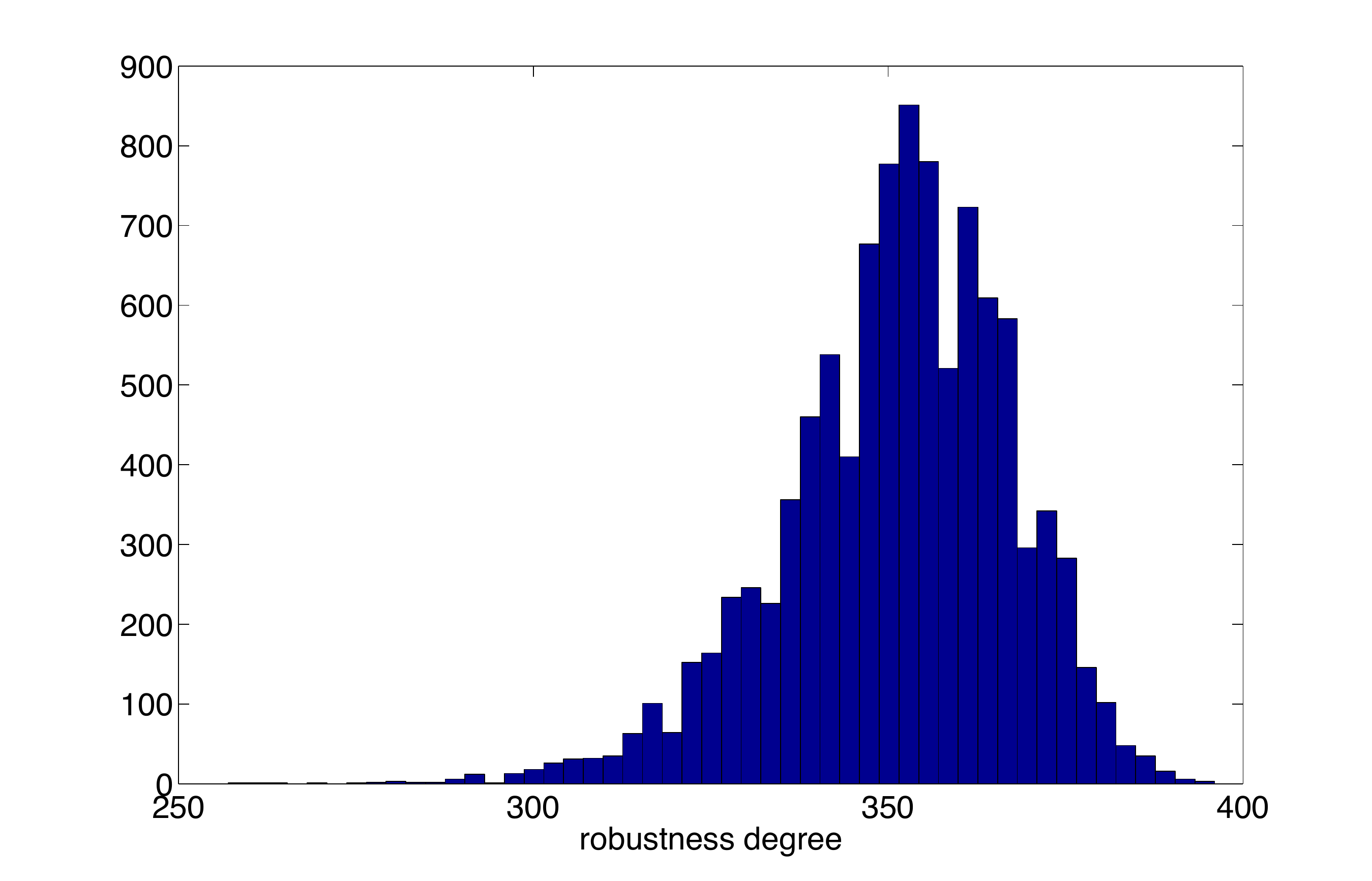} }
\end{center}
\caption{  The emulated robustness function in the optimisation of  $k_{3}$ (left). The distribution of the robustness score for $k_{3}=999.99$ (right).} %$k_{3}=965.02$. }
\label{fig:optimizationSchlog}
\end{figure}

\paragraph{Schl\"ogl system.}
%The optimization of the parameter $k_{1}$ gave a result in agreement with the  figure \ref{fig:k1}.
We set up the experiment as follows.
We combine  the robustness degree of the formula of Section \ref{sec:schlogl} and the satisfaction probability in the systems design problem asking, at the same time, to maximize the  robustness degree constraining the probability value to remain above  0.75. 
We varied  $k_{3}$ uniformly in  $[50, 1000]$, fixing  all  other parameters to the values of Table \ref{tab:schlogl}. We ran the GP-UCB optimisation algorithm by first  estimating the robustness degree and the satisfaction probability, using statistical model checking, for 30 points sampled randomly and uniformly from the parameter space, and then using the GP-UCB strategy to estimate the maximum of the upper bound function in a grid of 200 points. If in this grid a point is found with a larger value than those of the observation points, we compute the robustness and satisfaction probability also for this new point, and add it to the observations (thus changing the GP approximation). Termination happens when no improvement can be made after three grid resamplings. Further integration of local maximisation can further improve the method.
 \\  
In the experiment, repeated 10 times, we used a GP with radial basis kernel \cite{Bishop2006}, with length scale fixed to 0.5 (after standardisation of the parameter range to $[-1,1]$). The amplitude of the kernel was adaptively set to 60\%  of the difference between the maximum and the mean value of the robustness for the initial observations. The observation noise was experimentally fixed to 1, by monitoring the average standard deviation at different random parameter combinations.
\\
Results are shown in table \ref{tab:schlogl_result}. \change{As we can see, the result of the optimisation suggests that the more robust system satisfying the specification (i.e. remaining as much as possible above the threshold 300 for a sufficiently long amount of time) is the one obtained for $k_3 = 1000$. We can see that this is the case in Figure \ref{fig:robustnessSchlog}: the system becomes monostable, and $X$ stably remains above 550 units (corresponding to a robustness score above 250 with very high probability).}

%\begin{table}[!t]
%\begin{center}
%\begin{tabular}{|l|l|l|}
%\hline
%Parameter mean & Parameter range & Probability \\
%\hline
%$k_{1}=0.3$ $k_{3}=214.0679$ & [0.0299    0.0300] [147.3234  276.8391] & 1\\
%\hline
%Average Robustness &  Number of function evaluations &simulation time\\
%\hline
% 1250.4 & 11.4 & 243.49754 \\
% \hline
%\end{tabular}
%\end{center}
%\caption{The means of the results of ten experiments to optimize the parameters $k_{1}$in the range  $[5\cdot10^{-3},  3 \cdot 10^{-2}]$ and  and $k_{3}$   in the range $[100, 300]$.}
%\label{tab:schlogl_resultk1k3}
%\end{table}
%
%
%
%\begin{figure}
%\begin{center}
%\subfigure[The estimated robustness function]{\label{fig:final}
%\includegraphics[width=.485\textwidth]{figs/k1_obs.pdf} }
%\subfigure[The optimised distribution of the robustness score]{\label{fig:final}
%\includegraphics[width=.485\textwidth]{figs/schlogl_dist_k1_opt.pdf} }
%\end{center}
%\caption{ (a) the estimated robustness function evaluation (in black) for one of the 10 experiments to optimize the parameter $k_{1}$ (b) The distribution of the robustness score with $k_{1}=0.03$, it has probability degree 1 and average robustness   1250.4 }
%\label{fig:optimizationk1}
%\end{figure}

\begin{table}[!t]
\begin{center}
\begin{tabular}{|c|c|c|}
\hline
Parameter mean & Parameter range & Mean probability \\
\hline
$k_{3}=997.78$ &  [979.31  999.99] & 1\\
%$k_{3}=1000.00$ &  [1000.00  1000.00] & 1\\
\hline
Average Robustness & Number of function evaluations & Number of simulation runs\\
\hline
  348.97 & 34.4  & 3440 \\
%    350.67 & 17.4  & 1740 \\
 \hline
\end{tabular}
\end{center}
\caption{Statistics of the results of ten experiments to optimize the parameter $k_{3}$   in the range $[50, 1000]$.}
\label{tab:schlogl_result}
\end{table}

\paragraph{Repressilator.} We consider a different optimisation problem, in which we keep model parameters constant and we try to optimise the parameters of the formula to make the robustness score as large as possible. This can be seen as a sort of dual problem, in which the goal is to learn the emergent behaviour of the model in terms of the most robustly satisfied formula (of fixed structure).
Furthermore, the parametrisation of  a formula is usually an underestimated problem, as the satisfaction/robustness heavily depends on these parameters. This problem has been partially tacked e.g.\ in \cite{Rizk2008} for deterministic models, but never for stochastic ones, to authors' knowledge. 
In particular, we consider Formula (\ref{prop repress}) and  optimise the temporal delays $T_1$ in the range  $[0,500]$ and $T_2$ in the range $[1000, 7000]$. \change{This can be seen as an attempt to learn the best bounds on the oscillatory period, through the filter of the logical specification of oscillations of equation (\ref{prop repress}).}

\change{In this experiment, we used the same settings of the optimisation algorithms as for the Schl\'og\ system, save for the number of initial observations, set to 25 and constrained to lie in an equi-spaced grid. The parameters of the model are fixed to those shown in the caption of Figure \ref{fig:Rep}. }

In this case, due to the highly random duration of each oscillation cycle of the SHA model of the Repressilator, we expect to obtain a low value for $T_1$ and a large value for $T_2$. This is indeed the case (see Table \ref{tab:rep_result}) confirming the intrinsic instability of the oscillatory pattern of the model. Moreover, the large variability in the value of $T_1$ can indicate a scarce contribution of the parameter in the determination of the robustness.  We confirmed this intuition by emulating the average robustness score as a function of $T_1$ alone in the range $[0,500]$ for $T_2$ fixed to $6963$  (data not shown), obtaining an essentially flat function: the average robustness varies between 15.01 (near $T_{1}=0$) and 6.89 (near $T_{1}=500$). We expect that an heteroschedastic treatment of noise could improve this estimate of $T_1$.

\begin{table}[!t]
\begin{center}
\begin{tabular}{|c|c|c|}
\hline
Parameter mean & Parameter range & Mean probability \\
\hline
$T_{1}=231.5$
 $T_{2}=6993$ &  [0,  500] [6963, 7000] & 0.900 \\
\hline
Average Robustness & Number of function evaluations & Number of simulation runs\\
\hline
  17.20 &  35 & 3500  \\
  % Av 13.4560
 \hline
\end{tabular}
\end{center}
\caption{Statistics of the results of 10 experiments to optimize the parameter $T_{1}$   in the range $[0, 500]$ and the parameter $T_{2}$   in the range $[1000, 7000]$.}
\label{tab:rep_result}
\end{table}

%\begin{figure}
%\begin{center}
%\subfigure[The estimated robustness function]{\label{fig:final}
%\includegraphics[width=.485\textwidth]{figs/estimatek3in10_1000.pdf} }
%\subfigure[The optimised distribution of the robustness score]{\label{fig:final}
%\includegraphics[width=.485\textwidth]{figs/repress_dist_T2_opt.pdf} }
%\end{center}
%\caption{ (a) the estimated robustness function evaluation (in black) for one of the 10 experiments to optimize the parameter $k_{3}$, with kernel.RBFlengthScale = 0.5 and kernel.obsNoiseStDev = 4 (b) The distribution of the robustness score with $k_{3}=66.8954$ }
%\label{fig:optimization}
%\end{figure}

% !TEX root =  main.tex
\section{Conclusion}
\label{sec:conclusion}

\paragraph{Discussion.}  In this paper we investigate a notion of robustness of behaviours of stochastic models, extending the robustness degree of STL formulae in a probabilistic setting. Discussing two case studies, a bistable model and the Repressilator, we showed that the distribution of the robustness degree of a formula provides more information than the satisfaction probability alone, and its average can be used to  enforce robust behaviours by optimising it. Such optimisation is carried out using state-of-the-art optimisation algorithms coming from reinforcement learning, which emulate the true function from just few samples, and perform very well in a simulation based scenario. \change{Remarkably, the proposed approach to evaluate robustness and to system design  can be applied  both to CTMC and to SHA models.  We also briefly considered the problem of learning the most effective parameters of a given formula, in the sense of finding the parameter combination maximising the robustness score. This hints towards a more ambitious goal, that of finding machine learning procedures to learn the emergent behaviours (described as temporal logic formulae) from models and from experimental data. Many problems need to be faced to achieve this goal, like how to learn formula structure, how to avoid overfitting (with respect to both formula structure and parameters), how to deal with the curse of dimensionality afflicting GP-UCB and other optimisaton algorithms.}

% !TEX root =  hsb_2013.tex

\paragraph{Related Work.}
%\label{sec:related}
Temporal Logic (TL) is a very intuitive specification language to 
express formally the behavioural property emerging in a complex 
biological system. Several important extensions of TL, such as Metric Interval 
Temporal Logic (MITL) and Signal Temporal Logic (STL)~\cite{Maler2004}, have been 
introduced to deal with dense-time and real-valued signals, respectively.  

In the last years there was a great scientific effort to enrich the classic 
qualitative semantics of TL or \emph{satisfiability}
 (yes/no answer for the formula satisfaction of a trajectory) 
with more powerful and useful notions of quantitave 
semantics\cite{Fainekos2007a,Fainekos2009,Donze2010, Rizk2008,Annapureddy2011} 
(or \emph{robustness degree}), providing a real value measuring the level of satisfaction or 
violation for a trajectory of the property of interest. 

Several tools, such as BIOCHAM~\cite{biocham}, S-TaLiRo~\cite{Annapureddy2011} Breach~\cite{Donze2010a}, are now available 
to perform robustness analysis on the time series collected in wet-lab 
experiments or produced by simulation-based techniques. The robustness 
degree have been successfully employed in the analysis of ODE-based biological 
models, to tune the parameters that discriminates the behaviours observed 
experimentally (the so called design problem).

Donze et al. in~\cite{Donze2011} proposed a multi-step 
analysis, where they adopt STL to express dynamical properties and they use
robustness and sensitivity analysis to sample efficiently the parameter space, 
searching for feasible regions in which the model exhibit a particular 
behavior. A similar method appeared also in a previous paper~\cite{Donze2010b}
of one of the co-authors.

In~\cite{Bartocci2013} the authors proposed a new approach, based 
on robustness degree, for the design of a synthetic biological circuit whose 
behaviour is specified in terms of signal temporal logic (STL) formulae. Also in 
this case stocasticity was not taken into account.

For what concern the stochastic models, while the satisfiability analysis
has been considered as a discriminating criterion to tune the parameters
in the design process using both simulation-based statistical approximated 
methods~ \cite{lucaQEST13} and probabilistic exact methods~\cite{Bartocci2011}, 
to the best of our knowledge, we are not aware of approaches using the
robustness degree. Another related work in this sense is that of \cite{davidCAV}, where authors compute exactly upper and lower bounds on the satisfaction probability within a given region of the parameter space.

\paragraph{Future work.} The present work uses advanced machine learning concepts to address core problems in formal modelling; this is a relatively new line of work \cite{lucaQEST13} which opens significant new avenues for further research. From the practical point of view, more extensive testing and an efficient and robust implementation (exploiting some of the possible parallelisms e.g. in SMC) will be important for the tool to be adopted. From the theoretical perspective, we plan to use multi-objective optimisation to find good parametrisation for conflicting objectives. Another interesting direction is to combine the design problem with the inference problem, which has recently been addressed for a number of continuous time stochastic systems \cite{Ocone:hybrid13}; this would open the possibility of addressing the control problem for such systems, simultaneously inferring the state of the system and designing the optimal input to lead it to a desired state.

\paragraph{Acknowledgements.} Work partially supported by  the EU-FET project QUANTICOL (nr. 600708) and by FRA-UniTS.

\vspace{-0.3cm}

%\fi
\bibliographystyle{eptcs}
\bibliography{hsb}
\end{document}